\documentclass[twocolumn,aps,pra,showpacs,superscriptaddress,
]{revtex4}
\usepackage{graphicx}
\usepackage[varg]{txfonts}
\usepackage{dcolumn}
\usepackage{bm}
\usepackage{amssymb}
\usepackage{epsfig}
\usepackage[english]{babel}
\usepackage{latexsym}
\usepackage{graphics}
\usepackage{subfigure}

\def\be{\begin{equation}}
\def\ee{\end{equation}}
\def\bea{\begin{eqnarray}}
\def\eea{\end{eqnarray}}
\def\bse{\begin{subequations}}
\def\ese{\end{subequations}}

\begin{document}

\title{Initializing a Quantum Register from Mott Insulator States in Optical
Lattices}
\author{Chuanwei Zhang, V.W. Scarola, and S. Das Sarma}

\begin{abstract}
We propose and quantitatively develop two schemes to quickly and accurately
generate a stable initial configuration of neutral atoms in optical
microtraps by extraction from the Mott insulator state in optical lattices.
We show that thousands of atoms may be extracted and stored in the ground
states of optical microtrap arrays with one atom per trap in one operational
process demonstrating massive scalability. The failure probability during
extraction in the first scheme can be made sufficiently small ($\sim 10^{-4}$%
) to initialize a large scale quantum register with high fidelity. A
complementary faster scheme with more extracted atoms but lower fidelity is
also developed.
\end{abstract}

\affiliation{Condensed Matter Theory Center, Department of Physics, University of
Maryland, College Park, Maryland, 20742 USA}
\pacs{03.67.Lx}
\maketitle

\emph{Introduction:} Recently quantum computation using neutral atoms has
attracted much attention because neutral atoms are well isolated from the
environment and may offer a route to scalable quantum computation. Different
neutral atom quantum computing proposals \cite{Zoller,Mandel1,Calarco}
differ in trapping techniques. We focus on arrays of microscopic traps
(microtraps) that can be implemented using independent focused laser beams.
The typical size of a microtrap in current experiments is about $2\mu m$
because of the diffraction limit of a laser beam as well as experimentally
technical issues. Unlike an optical lattice, all microtraps can be moved
independently in position space, therefore they have advantages in single
atom addressability and controlled interactions between pairs of atoms. In
recent years, many schemes have been proposed for universal quantum
computation in microtrap systems \cite{Calarco}.

These schemes are based on the assumption that thousands of neutral atoms
can be prepared in the ground states of optical microtrap arrays with one
atom per trap, i.e. they assume the ability to initialize a neutral atom
quantum register. Although impressive experimental progress has been made in
trapping single atoms \cite{kurl}, such an assumption has not been fulfilled
because the trapping processes are random and not deterministic.
Furthermore, the trapped atoms are not in the trap ground states. These
difficulties have prevented neutral atom quantum computation architectures
from accomplishing the kind of impressive experimental progress recently
achieved in ion trap quantum computation \cite{Cirac1}. To overcome these
difficulties one proposal seeks to extract single atoms from a Bose-Einstein
condensate (BEC) by moving an optical dipole trap out of the condensate \cite%
{Diener}.

In this paper, we propose two concrete schemes to quickly extract thousands
of neutral atoms from the Mott insulator (MI) state in optical lattices \cite%
{Greiner} and prepare them in the ground states of optical microtrap arrays
with precisely one atom per trap. In the MI regime, atom number fluctuations
at each site are suppressed and a MI state may be obtained with proper
experimental parameters, as demonstrated in recent experiments \cite%
{Mandel1,Greiner,Folling}. Indeed, a perfect MI state with one atom per
lattice site may be implemented through purification processes \cite{Cirac2}%
. In MI states, atoms are isolated from each other and occupy the ground
states of each site making it difficult to realize a low density MI in long
lattice spacing ($d$) laser geometries \cite{Peil} where the MI energy
scales ($\sim d^{-2}$) fall below characteristic temperatures. On the other
hand, the size of a microtrap ($\sim 2\mu m$) is much larger than the
typical lattice spacing ($\sim 0.4\mu m$), which make it difficult to
extract atoms to microtraps directly from a MI state. Therefore the main
challenge, and also the goal of this paper, is to transfer single atoms from
optical lattices with a \emph{short} lattice period (SPOL) to spatially
separate optical microtraps.

We show in the first scheme that such transfer process may be accomplished
using hyperfine state dependent optical lattices with a \emph{long} lattice
period (LPOL), microwave radiation, and resonant \textquotedblleft
removing\textquotedblright\ lasers. A LPOL is created by intersecting two
laser beams at a certain angle, as demonstrated in a recent experiment \cite%
{Peil}. The LPOL induces position-dependent energy shifts of the hyperfine
states of atoms, which, when combined with microwave radiation and resonant
\textquotedblleft removing\textquotedblright\ lasers, expel many atoms out
of the optical lattice and form a patterned loaded optical lattice with one
atom per $n$ lattice sites ($n\geq 3$ is an integer). The remaining atoms in
the lattices are well separated, and can be adiabatically transferred to the
ground states of optical microtrap arrays with one atom per trap. With our
scheme, thousands of atoms may be extracted and the failure probability in
extraction can be kept below $\sim 10^{-4}$. In the second scheme, the
transfer process is accomplished using state-dependent focused lasers
without expelling atoms out of the optical lattice, therefore more atoms can
be extracted but the failure probability is much higher. These complementary
schemes, one with very high fidelity (but relatively lower speed) and the
other with very fast speed (but relatively lower fidelity), take advantage
of the robust localization inherent in the MI to isolate atoms. The whole
operational process is simple and within currently accessible experimental
technology.

The first extraction scheme includes four operational steps and their time
sequences are schematically plotted in Fig. 1a. In the following, we explain
each step and study various errors that could lead to failures in
extraction. For simplicity, we focus on a one dimensional geometry but we
emphasize that our technique can be straightforwardly applied to two
dimensional arrays.

\emph{Step (I) Initial Mott state:} Consider a pure $^{87}$Rb BEC prepared
in the hyperfine ground state $\left\vert 0\right\rangle \equiv \left\vert
F=1,m_{F}=-1\right\rangle $ and confined in a quasi-one dimensional ($x$
direction) harmonic magnetic trap. Along the transverse direction, the
atomic dynamics are frozen out by high frequency optical traps \cite%
{Meyrath1}. An optical lattice along the $x$ direction with wavelength $%
\lambda _{s}=850nm$ is ramped up adiabatically ($\sim 200ms$) to a large
potential depth of $V_{s}=50E_{R}$ such that the BEC is converted into a MI
state with roughly one atom per lattice site, using properly chosen trapping
parameters and number of atoms \cite{Greiner}. Here $E_{R}=h^{2}/2m\lambda
_{s}^{2}$ denotes the recoil energy. This MI state may have defects \cite%
{Folling} and clever purification schemes \cite{Cirac2} may yield a nearly
perfect MI state with exact one atom per lattice site. Such a perfect MI
state is the reservoir for single atom extraction and provides the starting
point of our scheme.

\begin{figure}[t]
\begin{center}
\resizebox*{8cm}{5cm}{\includegraphics*{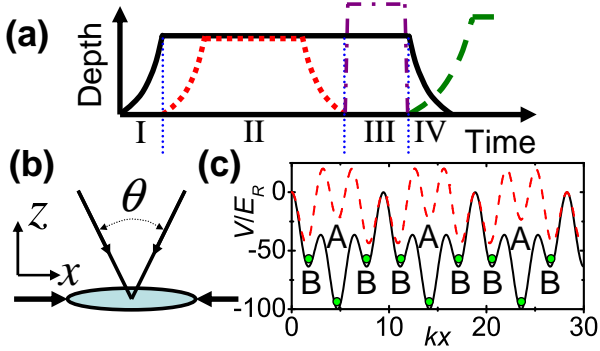}}
\end{center}
\par
\vspace*{-0.5cm}
\caption{(color online). Schematic plot of the single atom extraction. (a)
Time sequence for laser depths at four extraction steps. Solid line (SPOL),
dotted line (LPOL), dashed dotted line (removing laser), dashed line
(microtrap laser). (b) Geometry of additional lasers defining the LPOL. (c)
Site specific energies in the optical lattice. Solid and dashed lines
correpsond to the optical potentials for atoms at states $\left\vert
0\right\rangle $ and $\left\vert 1\right\rangle $ respectively. $A$ labels
target atoms for extraction and $B$ labels other atoms.}
\label{rr}
\end{figure}

\emph{Step (II) Selective Depopulation}: Here atoms at specific, unwanted
sites are transferred to another hyperfine state. Two $\sigma ^{+}$%
-polarized laser beams intersecting at an angle $\theta $ (Fig.1b), which
drive the $5S\rightarrow 5P$ transition and has a detuning $\Delta
_{2}=-2\pi \times 3608GHz$ to the $5^{2}P_{3/2}$ state (corresponding to a
wavelength $\lambda _{l}=787.6nm$), are adiabatically ramped up. These two
beams form a LPOL along the $x$ direction with the lattice period $\eta
_{l}=\lambda _{l}/\left[ 2\sin \left( \theta /2\right) \right] $. Here we
require $\eta _{l}$ to be $n$ times that of the short wavelength optical
lattice period $\lambda _{s}/2$, that is, $\theta =2\arcsin \left[ \lambda
_{l}/n\lambda _{s}\right] $. The LPOL induces energy shifts 
\begin{equation}
\delta E_{\pm }\left( r\right) =\frac{3\pi c^{2}I\left( r\right) }{2}%
\sum_{q=1,2}\frac{\Gamma _{q}\left\vert c_{\pm q}\right\vert ^{2}}{\omega
_{q}^{3}\Delta _{q}}  \label{shift1}
\end{equation}%
for two finestructure ground states $\left\vert \pm \right\rangle
=\left\vert 5S:j=1/2,m_{j}=\pm 1/2\right\rangle $, where $I\left( r\right) $
is the intensity of the laser, $\Gamma _{q}$ ($q=1,2$) is the decay rate for
states $5^{2}P_{1/2}$ and $5^{2}P_{3/2}$. $\omega _{q}$ ($\Delta _{q}$) is
the frequency (detuning) for the transition from $5S$ to $5^{2}P_{1/2}$ and $%
5^{2}P_{3/2}$. $c_{+1}=0$, $c_{+2}=1$, $c_{-1}=-\sqrt{2/3}$, $c_{-2}=\sqrt{%
1/3}$ are transition coefficients. Two hyperfine states $\left\vert
0\right\rangle $ and $\left\vert 1\right\rangle \equiv \left\vert
F=2,m=-2\right\rangle $ can be written as $\left\vert 0\right\rangle
=1/4\left\vert -\right\rangle +3/4\left\vert +\right\rangle $, $\left\vert
1\right\rangle =\left\vert -\right\rangle $ using the respective
Clebsch-Gordan coefficients. Therefore the energy shifts for states $%
\left\vert 0\right\rangle $ and $\left\vert 1\right\rangle $ are $\delta
E_{0}=\delta E_{-}/4+3\delta E_{+}/4$, and $\delta E_{1}=\delta E_{-}$,
which give the total shift of the hyperfine splitting between $\left\vert
0\right\rangle $ and $\left\vert 1\right\rangle $, $\delta E=\delta
E_{1}-\delta E_{0}=\frac{\alpha _{1}}{2\Delta _{1}}-\frac{\alpha _{2}}{%
2\Delta _{2}}$, where $\alpha _{q}=3\pi c^{2}\Gamma _{q}I/2\omega _{q}^{3}$.
Following a similar procedure, we find the spontaneous emission rates for
atoms at $\left\vert 0\right\rangle $ and $\left\vert 1\right\rangle $ are%
\begin{equation}
\gamma _{0}=\frac{\alpha _{1}\Gamma _{1}}{6\Delta _{1}^{2}}+\frac{5\alpha
_{2}\Gamma _{2}}{6\Delta _{2}^{2}},\text{ }\gamma _{1}=\frac{2\alpha
_{1}\Gamma _{1}}{3\Delta _{1}^{2}}+\frac{\alpha _{2}\Gamma _{2}}{3\Delta
_{2}^{2}}.  \label{rate2}
\end{equation}%
Notice that the wavelength $\lambda _{l}=787.6nm$ (i.e. the detuning $\Delta
_{2}=-2\pi \times 3608GHz$) is optimized to obtain the maximal ratio $\eta
=\delta E/\gamma $ between the shift $\delta E$ and the rate $\gamma =\max
\left\{ \gamma _{0}\text{, }\gamma _{1}\right\} $.

The energy shifts induced by the LPOL are spatially dependent for atoms
confined in the SPOL. In Fig. 1c, we plot the combined lattice potentials
for atoms at $\left\vert 0\right\rangle $ and $\left\vert 1\right\rangle $
with $n=3$. We see that the LPOL induces two different shifts of the
hyperfine splitting, depending on the positions of atoms ($A$ or $B$), and
the difference $\delta =\delta E\left( A\right) -\delta E\left( B\right) $
can be adjusted and chosen to be $\delta =52E_{R}$. Applying the adiabatic
condition, we estimate the ramp up time to be $44\mu s$ that corresponds to
a $10^{-4}$ probability for excitation to higher bands.

We then apply a microwave $\pi $ pulse to flip the quantum states of atoms
at position $B$ (Fig. 1c) from $\left\vert 0\right\rangle $ to $\left\vert
1\right\rangle $. The microwave is resonant with the hyperfine splitting of
atoms at $B$, but has a detuning $\delta $ for atoms at $A$. A
time-dependent microwave pulse with frequency $\Omega \left( t\right)
=\Omega _{0}\exp \left( -\omega _{0}^{2}t^{2}\right) $ ($-t_{f}\leq t\leq
t_{f}$) is used to perform the $\pi $ pulse. For a set of microwave
parameters $\omega _{0}=\delta /4=13E_{R}/\hbar $, $t_{f}=5/\omega _{0}$,
and $\Omega _{0}=\pi /\left[ \int_{-t_{f}}^{t_{f}}\exp \left( -\omega
_{0}^{2}t^{2}\right) dt\right] \approx 23E_{R}/\hbar $, the pulse flips the
quantum state of atom $B$ from $\left\vert 0\right\rangle $ to $\left\vert
1\right\rangle $ in $2t_{f}=38.6\mu s$, while the error to flip atom $A$ to
state $\left\vert 1\right\rangle $ is found to be $5.9\times 10^{-6}$ by
numerically integrating the Rabi equation \cite{Metcalf} that describes the
coupling between $\left\vert 0\right\rangle $ and $\left\vert 1\right\rangle 
$ through the microwave pulse. The LPOL are adiabatically turned off after
the microwave pulse. During the whole process time $\tau $, the probability
for spontaneous scattering of one photon from each atom is estimated to be $%
P=\int_{0}^{\tau }\gamma dt\approx 1\times 10^{-4}$.

\emph{Step (III) Remove Non-target Atoms}: In this step, atoms at $B$ are
removed from the trap by applying a $\sigma ^{-}$-polarized
\textquotedblleft removing\textquotedblright\ laser that drives a resonant
cycling transition $\left\vert 1\right\rangle \rightarrow \left\vert
2\right\rangle \equiv \left\vert 5^{2}P_{3/2}:F=3,m_{F}=-3\right\rangle $.
Scattering photons from the laser push (or heat) non-target atoms $B$ at $%
\left\vert 1\right\rangle $ out of the trap without affecting target atoms $A
$ at state $\left\vert 0\right\rangle $ because of the large hyperfine
splitting ($\nu \approx 2\pi \times 6.8GHz$) between the two states. To push
(or heat) atoms out of a trap with depth $U_{0}$, the number of spontaneous
emission photons needs to be at least $n_{p}=U_{0}/2E_{R}$ \cite{kurl}. For
instance, $25$ photons are needed for each non-target atom to remove it from
an optical lattice with depth $U_{0}=50E_{R}$. The dynamics of the photon
scattering process are described by the optical Bloch equation \cite{Metcalf}%
, from which we can numerically calculate the number of scattering photons $%
n_{p}$ for both target and non-target atoms. We find the number of
scattering photons can reach $25$ in a short period $\sim 1\mu s$ for atoms $%
B$, but it is only $\sim 10^{-5}$ for atoms $A$. Therefore the impact of the
resonant laser on the target atoms $A$ can be neglected. Note that hot
non-target atoms tunnel more easily in the optical lattice, which enhances
the collision probability between target and non-target atoms at different
lattice sites. However, because of the high lattice depths ($\sim 50E_{R}$),
the tunneling rate is quite low ($\sim 0.01E_{R}$) on average for hot
non-target atoms, which yields a long tunneling time ($\sim 100ms$). The
short lifetime ($\sim 1\mu s$) of hot non-target atoms makes the collision
probability very small ($\sim 1\mu s/100ms=10^{-5}$). The effect of
interatomic collisions on the target atoms can therefore be neglected.

$\emph{Step}$ $\emph{(IV)}$ $\emph{Transfer}$ $\emph{to}$ $\emph{Microtraps}$%
: In this last step, the remaining target atoms $A$ at the optical lattice
sites are adiabatically distributed to the ground state of optical microtrap
arrays with one atom per trap. The optical microtraps may be focused far
red-detuned lasers or high frequency blue-detuned optical traps using
Hermite-Gaussian TEM$_{01}$ mode beams \cite{Meyrath1}. Each optical
microtrap is focused near one target atom and contains only one atom because
of the large spacing between atoms in the lattice. In the distribution
process, the effective potentials the target atoms experience may be
approximated as harmonic potentials with the trapping frequency $\varpi
\left( t\right) =\sqrt{\left( 4V_{f}\left( t\right) /w^{2}+2V_{L}\left(
t\right) k^{2}\right) /m}$, where $V_{f}\left( t\right) $ and $V_{L}\left(
t\right) $ are the potential depths of the microtraps and the optical
lattice respectively, $w$ is the beam waist of the microtrap lasers. $%
V_{f}\left( t\right) $ and $V_{L}\left( t\right) $ may be varied
simultaneously to adjust $\varpi \left( t\right) $ from its initial value $%
\varpi \left( 0\right) =(2E_{R}/\hbar )\sqrt{V_{L}\left( 0\right) /E_{R}}$
to the final expected $\varpi \left( t_{0}\right) $. If $\varpi \left(
t_{0}\right) =\varpi \left( 0\right) $, the final microtrap potential depth $%
V_{f}\left( t_{0}\right) =V_{L}\left( 0\right) k^{2}w^{2}/2\approx 1366E_{R}$
($\sim 104\mu K$) for initial parameters $V_{s}\left( 0\right) =50E_{R}$,
and $w=1\mu m$.

When a deeper or shallower microtrap potential depth is needed, the trapping
frequency $\varpi \left( t\right) $ should be adjusted so that the adiabatic
condition is satisfied%
\begin{equation}
\hbar \left\vert \frac{d\varpi \left( t\right) }{dt}\right\vert =\xi \frac{%
\left( \Delta E_{g}\right) ^{2}}{\left\vert \left\langle \phi
_{e}\right\vert \frac{\partial H}{\partial \varpi }\left\vert \phi
_{g}\right\rangle \right\vert },  \label{adab}
\end{equation}%
where $\left\vert \phi _{g}\right\rangle $, and $\left\vert \phi
_{e}\right\rangle $ are the ground and the excited state wavefunction, $%
\Delta E_{g}$ is the energy gap between two states, and $\xi $ is the
adiabaticity parameter. Because of the parity of the wavefunctions, the
lowest possible excitation is to the second excited state, which gives $%
\Delta E_{g}=2\hbar \varpi $, and $\left\langle \phi _{e}\right\vert \frac{%
\partial H}{\partial \varpi }\left\vert \phi _{g}\right\rangle =$ $\hbar /%
\sqrt{2}$. The adiabatic condition, Eq. (\ref{adab}), yields a
time-dependent trapping frequency $\varpi \left( t\right) =\varpi \left(
0\right) /\left( 1\mp 4\sqrt{2}\xi \varpi \left( 0\right) t\right) $, where $%
\mp $ correspond to a deeper and shallower final trapping frequency
respectively.

\begin{figure}[t]
\includegraphics[scale=0.4]{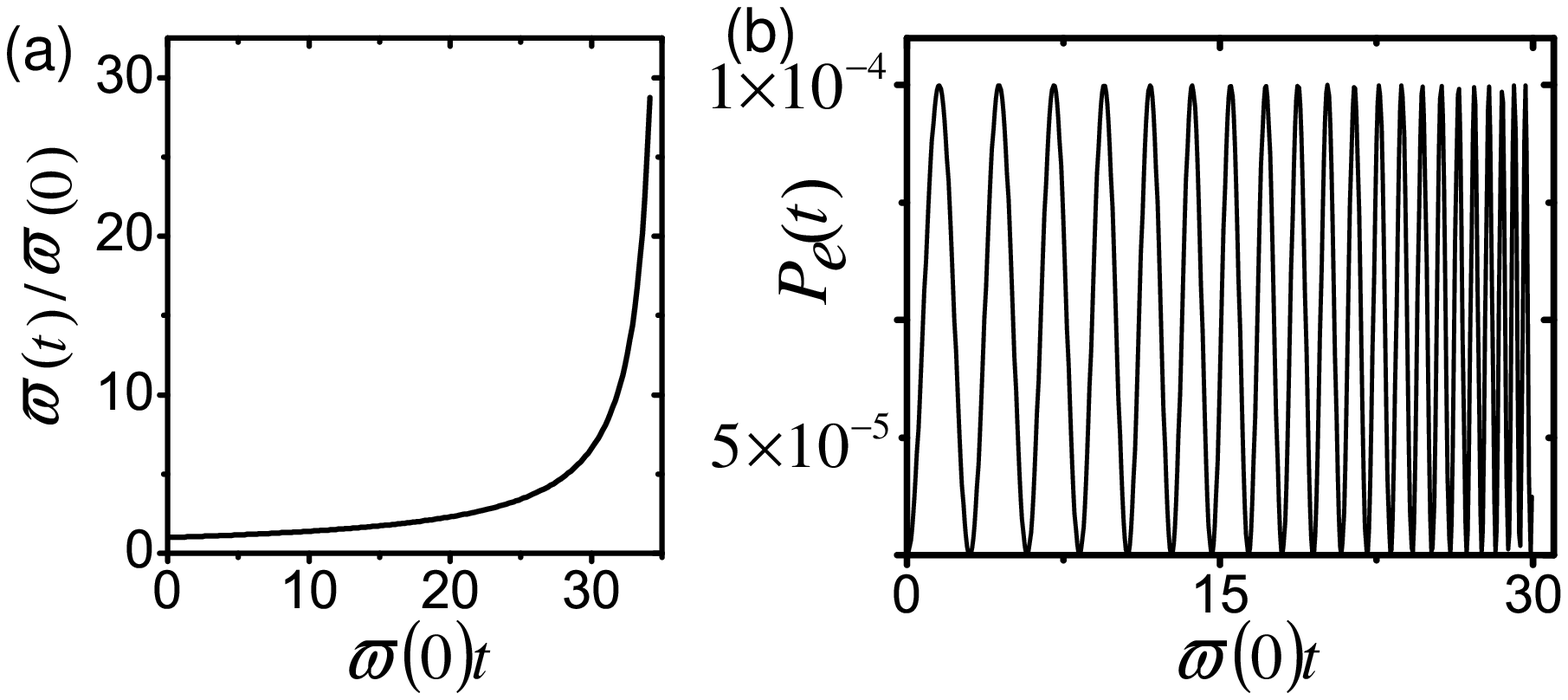}
\caption{The adiabatic process for the distribution of the target atoms from
the optical lattice to arrays of optical dipole microtraps with higher
trapping frequencies. Adiabaticity parameter is $\protect\xi =0.005$. (a)
Trapping frequency with respect to time. (b) The probability $P_{e}\left(
t\right) =\left\vert c_{e}\right\vert ^{2}$ for atoms to be at the second
excited state of the effective harmonic trap with respect to time. }
\label{rr4}
\end{figure}

Under the adiabatic approximation, the quantum states of the target atoms
can be expanded using the time-dependent basis $\varphi \left( t\right)
=c_{g}\left( t\right) \phi _{g}\left( \varpi \left( t\right) \right)
+c_{e}\left( t\right) \phi _{e}\left( \varpi \left( t\right) \right) $,
where $\phi _{g}\left( \varpi \left( t\right) \right) $, and $\phi
_{e}\left( \varpi \left( t\right) \right) $ are the adiabatic ground and
second excited states of the Hamiltonian $H\left( \varpi \left( t\right)
\right) $ with the associated eigenenergies $E_{g}=\hbar \varpi \left(
t\right) /2$ and $E_{e}=5\hbar \varpi \left( t\right) /2$. Inserting this
expansion into the the Schr\"{o}dinger equation for a single target atom $A$
yields a coupled equation for the coefficients $c_{g}\left( t\right) $ and $%
c_{e}\left( t\right) $: 
\begin{equation}
i\hbar \frac{d}{dt}\left( 
\begin{array}{c}
c_{g} \\ 
c_{e}%
\end{array}%
\right) =\left( 
\begin{array}{cc}
E_{g} & \varkappa \left( t\right) \\ 
-\varkappa \left( t\right) & E_{e}%
\end{array}%
\right) \left( 
\begin{array}{c}
c_{g} \\ 
c_{e}%
\end{array}%
\right) ,  \label{coup}
\end{equation}%
where $\varkappa \left( t\right) =-i\hbar \frac{d\varpi \left( t\right) }{dt}%
\frac{\left\langle \phi _{e}\right\vert \frac{\partial H}{\partial \varpi }%
\left\vert \phi _{g}\right\rangle }{\Delta E_{g}}=i\xi \Delta E_{g}$. This
equation can be solved analytically to give the occupation probability at
the second excited states 
\begin{equation}
P_{e}\left( t\right) =4\xi ^{2}\sin ^{2}\left( \ln \left( 1\mp 4\sqrt{2}\xi
\varpi \left( 0\right) t\right) /4\sqrt{2}\xi \right) .  \label{prob}
\end{equation}

In Fig. 2, we plot the trapping frequency $\varpi \left( t\right) $ and
excitation probability $P_{e}\left( t\right) $ with respect to time for a
deeper final trap. We see $P_{e}\left( t\right) $ oscillates with time, but
has the maximum $4\xi ^{2}$. For adiabaticity parameter $\xi =0.005$, the
maximal excitation probability is $10^{-4}$. The total distribution time is $%
T=\left( 1-\varpi \left( 0\right) /\varpi \left( T\right) \right) /(4\sqrt{2}%
\xi \varpi \left( 0\right) )$. For instance, $T\approx 94\mu s$ is needed to
transfer target atoms from the optical lattice to microtraps with trapping
frequency $\varpi \left( T\right) =4\varpi \left( 0\right) $ (corresponding
to a $1.66$ $mK$ trap potential) with $10^{-4}$ excitation probability. We
note that this transfer time is much shorter than the characteristic hopping
time ($\sim 5s$ for $V_{s}=50E_{R}$) of atoms within the depleted optical
lattice.

Combining all four operational steps, we find that thousands of atoms can be
extracted from the optical lattice to microtraps in less than $300\mu s$.
For instance, in a one dimensional optical lattice with 300 usable atoms for
single atom extraction, we can extract 100 atoms (one per three lattice
sites). In a two dimensional trap, the same process can extract 1/9 of total
atoms ($\sim 10^{4}$) simultaneously, which offers a large scale up in the
initialization stage of a quantum computation using microtrap arrays.

The failure probability of extraction does not decrease with an increasing $n
$. That is because the microtrap potential is already very weak at a $\sim
1.3\mu m$ displacement (the position of the neighboring atoms in a $n=3$
superlattice) from the trap center for a typical $\sim 2\mu m$ microtrap.
The potential has a negligible effect on the process of transferring
neighboring atoms to other microtraps. A large $n$ lattice does not
contribute to the largest error source of the scheme: the heating due to
spontaneously scattered photons from atoms in the state-dependent lattice in
step (II).

\emph{A Speedup Scheme: }In the above scheme, the qubit supply time for
quantum computation is limited by the period for preparing Bose-Einstein
condensates, which is typically on the order of minutes. Notice that in 
\emph{step (III)} most atoms (2/3 in 1D or 8/9 in 2D lattices) are lost from
the lattice by applying the resonant "removing" laser. In the following, we
propose a scheme for single atom extraction without removing non-target
atoms, which can then be recycled for further extraction processes. In this
scheme, $87\%$ of the atoms can be extracted within several
superfluid-insulator transition cycles ($\sim 1s$), considerably speeding up
the rate for supplying fresh qubits for a neutral atom quantum computer.
However, the scheme induces much larger spontaneous emission probability of
photons and excitation probability to high motional states for atoms, which
significantly degrades the fidelity of the initial qubit, therefore further
cooling is needed to obtain qubits with high fidelity.

\begin{figure}[t]
\includegraphics[scale=0.6]{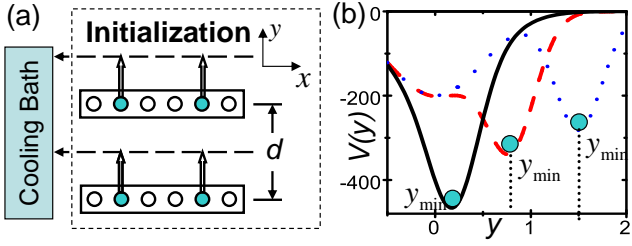}
\caption{(color online) (a) Schematic plot for the single atom extraction
from paralell one dimensional lattices. The distance $d=5\protect\mu m$.
Filled and opened circles correspond to target and non-target atoms
respectively. (b) Optical potentials for target atoms along the $y$
direction. The length and energy units are $\protect\sigma _{c}$ and $\hbar
^{2}/2m\protect\sigma _{c}^{2}$ respectively. $a\left( t\right) =0.2\protect%
\sigma _{c}$ (solid line); $a\left( t\right) =0.8\protect\sigma _{c}$
(dashed line); $a\left( t\right) =1.5\protect\sigma _{c}$ (dotted line).}
\label{rr5}
\end{figure}

A schematic for the speed up scheme is plotted in Fig. 3a. Consider a series
of parallel well separated ($d=5\mu m$) 1D optical lattices (along the $x$
direction) in the $xy$ plane. In the speedup scheme, $\emph{s}$\emph{teps (I)%
}-\emph{(II) }are still preformed to place all target atoms for extraction
into state $\left\vert 0\right\rangle $ and non-target atoms to $\left\vert
1\right\rangle $. We then ramp up a focused laser with the detuning $\Delta
_{0}=-2\pi \times 780GHz$ from the $5S\rightarrow 6^{2}P_{3/2}$ transition
so that the focused lasers only induce red-detuned traps for target atoms at
state $\left\vert 0\right\rangle $, but do not affect non-target atoms at
state $\left\vert 1\right\rangle $ \cite{Zhang}. The focused lasers
adiabatically move along $y$ direction to take target atoms out of the
optical lattices, without affecting non-target atoms at state $\left\vert
1\right\rangle $ (Fig.3a). The atoms inside the focused lasers are then
adiabatically transferred to far-detuned dipole traps to suppress
spontaneous emission of photons. These dipole traps are moved along the $x$
direction to a cooling bath to improve the fidelity of the initial qubits.
At the same time, the Mott insulator states in the optical lattices are
melted by adiabatically ramping down the depth of the optical lattice. The
trap parameters are adjusted so that another Mott state with one atom per
lattice site can be obtained as the lattice depths are adiabatically ramped
up. Repeating the above steps, we can extract another 1/3 of the remaining
atoms. After 5 such cycles, about $1-(2/3)^{5}\approx 87\%$ atoms can be
extracted. Finally, the remaining atoms inside the optical lattices are
discarded and a new BEC must be produced to continue the process.

To extract the atoms adiabatically from optical lattices avoiding
excitations to higher bands of the focused laser traps, high potential
depths are needed, which leads to high spontaneous photon scattering
probabilities. Therefore a good strategy balances these two sources for the
degradation of initial qubit fidelity. Assume that the optical potentials
along the $y$ direction for target atoms are 
\begin{eqnarray}
V\left( y\right) &=&-V_{c}\exp \left( -2y^{2}/\sigma _{c}^{2}\right)
\label{potential} \\
&&-V_{f}\exp \left( -2\left( y-a\left( t\right) \right) ^{2}/\sigma
_{f}^{2}\right) ,  \nonumber
\end{eqnarray}%
where $V_{c}$ and $V_{f}$ are the potential depths for the 1D confinement
and the focused laser, respectively. Gaussian beam approximations with
waists $\sigma _{c}$ and $\sigma _{f}$ have been used for a rough estimate. $%
a\left( t\right) $ is the position of the focused laser center and its rate
of change should satisfy the adiabatic condition $\hbar \left\vert da\left(
t\right) /dt\right\vert =\bar{\xi}\Delta E_{g}^{2}/\left\vert \left\langle
\phi _{e}\right\vert \partial H/\partial a\left\vert \phi _{g}\right\rangle
\right\vert $, where the energy gap $\Delta E_{g}$ and transition matrix
element $\left\vert \left\langle \phi _{e}\right\vert \partial H/\partial
a\left\vert \phi _{g}\right\rangle \right\vert $ can be evaluated for
different potentials $V\left( a\left( t\right) \right) $ by solving the
single particle Schr\"{o}dinger equation. The potential minimum position $%
y_{\min }$ for the target atom is determined through $\partial V/\partial
y=0 $. The wavefunction of the target atom is expanded in a harmonic
oscillator basis $\Psi _{n}\left( y\right) =\left( \varkappa /\pi \right)
^{1/4}\exp \left( -\varkappa \left( y-y_{\min }\right) ^{2}/2\right)
H_{n}\left( \sqrt{\varkappa }y\right) /\sqrt{2^{n}n!}$ around the potential
minimum $y_{\min }$, where the oscillation frequency $\varkappa =\left(
\partial ^{2}V/\partial y^{2}\right) _{y\min }/m$ and $H_{n}\left( \sqrt{%
\varkappa }y\right) $ is the Hermite polynomial. The Hamiltonian for the
target atoms evaluated in this basis for different $a\left( t\right) $ gives
the matrix representation $H_{nm}\left( a\left( t\right) \right)
=\left\langle \Psi _{n}\right\vert H\left\vert \Psi _{m}\right\rangle $ ($%
n,m\leq 10$ is enough for our calculation). Diagonalization of the matrix
yields the eigenenergies and eigenfunctions, which determine $\Delta E_{g}$
and $\left\vert \left\langle \phi _{e}\right\vert \partial H/\partial
a\left\vert \phi _{g}\right\rangle \right\vert $ for different $a\left(
t\right) $. The total moving time can be estimated using 
\begin{equation}
T=\int_{0}^{a_{f}}\hbar \left\vert \left\langle \phi _{e}\right\vert
\partial H/\partial a\left\vert \phi _{g}\right\rangle \right\vert /\bar{\xi}%
\Delta E_{g}^{2}da  \label{time}
\end{equation}%
where $a_{f}$ is the final position of the focused laser.

In Fig.3b, we plot the optical potential $V\left( y\right) $ for a set of
parameters $\sigma _{c}\approx 0.93\mu m$, $\sigma _{f}\approx 0.46\mu m$, $%
V_{c}=200\hbar ^{2}/m\sigma _{c}^{2}$, $V_{f}=280\hbar ^{2}/m\sigma _{c}^{2}$
and three different $a\left( t\right) $. Applying the above procedure with
these parameters, we estimate the extraction time $\sim 5ms$ and the
excitation probability to high bands $\sim 7\times 10^{-3}$, with the
spontaneous scattering probability $\sim 10^{-2}$. We see that the fidelity
of the atoms is not as high as that for the original scheme. Therefore
further cooling is needed to significantly improve the fidelity of initial
qubits.

\emph{Conclusion:} We propose two schemes for extracting thousands of atoms
simultaneously from the Mott insulator state in optical lattices to optical
microtrap arrays with one atom per trap. The extracted atoms stay at the
ground states of the microtraps. In the first scheme, about $11\%$ of atoms
from a BEC can be extracted and the failure probability is $\sim 10^{-4}$
with properly chosen experimental parameters. In the speedup scheme, about $%
87\%$ of atoms from a BEC can be extracted, but the failure probability is
much higher $\sim 10^{-2}$. We provide a detailed quantitative analysis
validating the feasibility of our proposed schemes for neutral atom quantum
computation.

We thank S.L. Rolston for valuable discussion. This work is supported by
ARO-DTO, ARO-LPS, LPS-NSA.

\end{document}